# Novel two-phase method for supercritical water flow

Piyush Mani Tripathi, Saptarshi Basu

Department of Mechanical Engineering, Indian Institute of Science, Bangalore 560012 India

## ABSTRACT

The current resurgence in the phase diagram study beyond the critical point has questioned the conventional belief of supercritical fluid as a single phase with varying properties. On the same line, a novel two-phase approach has been proposed to study the supercritical flow with heat transfer deterioration (HTD) phenomena. The Volume of Fluid (VOF) multiphase model has been used to analyze the flow, and the simulation result reasonably predicts the wall temperature peaks. Moreover, the velocity and turbulent kinetic energy profiles for different axial locations explain the occurrence of HTD, and it is on par with the pre-existing numerical method. Besides, the quantitative analysis presented in the paper expounds on qualitative understanding. The parametric study of the thermophysical properties revealed that the density variation is the primary cause of HTD in supercritical flows. So banking on this conclusion, the propounded study focuses on the forces generated due to the density variation. This technique equips us with a holistic picture of the supercritical flows. It leads to the inference that for no HTD effects to be present, buoyancy and inertia forces have to be of comparable magnitude throughout the flow. In addition, the pseudo phase change model outclasses the existing research by rendering us the volume fraction data. Mapping of this variable reveals a sudden jump in the lighter phase's thickness near the wall at the site of HTD, which is also reflected as a maximum in the nondimensional two-phase thickness (P) plot. Nevertheless, this observation is only restricted to HTD caused by buoyancy. One can draw a similarity of this occurrence to a phenomenon of film boiling in subcritical fluids. In the end, a theoretical expression has been conceptualized for computing the two-phase thickness (h) value, which can serve as a fundamental length scale in supercritical flows as it marks the region of highest property gradient near the wall.

## 1. Introduction

The vanishing of phase transition above the critical point has fascinated researchers right from its inception. Several studies conducted at the supercritical pressure (higher than critical value) have shown a drastic variation in the fluid's thermophysical property across a very narrow region. This region is generally known as the pseudo-critical line or widom line emanating from the critical point into the supercritical domain. The widom line can be defined in various ways, such as the locus of state with maximum response functions (like specific heat) or the line joining the density variation's inflection point for a given pressure.

| Nomenclature | | | |
|---|---|---|---|
| x | Axial distance from the inlet (m) | H | Specific enthalpy (kJ/kg) |
| r | Radial distance from the axis (mm) | HTD | Heat transfer deterioration |
| dx | Axial grid size (mm) | Pe | Peclet number |
| dr | Radial grid size (mm) | Pr | Prandtl number |
| $y^+$ | Nondimensional wall distance | P | Dimensionless two-phase thickness |
| D, R | Diameter and radius (mm) | VOF | Volume of fluid |
| Re | Reynold number | Greek letters | |
| S | Entropy (J/kgK) | $\rho$ | Density (kg/m$^3$) |
| U | Average velocity (m/s) | $\mu$ | Viscosity (Pa.s) |
| u,v | Axial and radial velocity (m/s) | $\beta$ | Volume fraction |
| T | Temperature (K) | | |
| q | Heat flux (kW/m$^2$) | Subscripts and superscripts | |
| G | Mass flux (kg/s) | in | Inlet |
| K | Thermal conductivity (W/mK) | out | Outlet |
| $C_p$ | Specific heat capacity (J/kgK) | w | Wall |
| $\dot{m}$ | Mass transfer across phases (kg/s/m$^3$) | pc | Pseudocritical |
| TKE (k) | Turbulent Kinetic energy (m$^2$/s$^2$) | b | Bulk |
| Avg. | Average of numbers | t,e | Turbulent and effective |
| p | Pressure (Pa) | l,v | Liquid and vapor |
| F | Forces (N) | sat | saturation |
| C.A | Cross-sectional area (m$^2$) | a, b | Acceleration and buoyancy |
| SA | Surface area (m$^2$) | th, nu | Theoretical and numerical |
| h | Two-phase thickness (mm) | co, an | Core and annular |

Interestingly, this pseudo-critical line demarcates the fluid between the liquid-like and vapor-like behaviour as shown in Figure 1[1]. In other words, from the macroscopic point of view, it may seem that the density is continuous, but there exist non-homogeneity at the molecular level. According to Hall [2], with an increase in temperature (at supercritical pressure), liquid alters into a liquid-like mass distributed in a vapor phase, and eventually, these liquid bundle's size decreases. As a consequence, the whole structure looks like a vapor with a high degree of interaction. Although this understanding existed long ago, supercritical fluid has always been perceived as a single phase with property variation to understand the flow behavior.

But the recent works like [3,4] on the transition of states across the pseudo-critical line have challenged the traditional belief that supercritical fluid consists of a single phase. Banauti [3] has tried to demonstrate the prevalence of pseudo boiling phenomena upon crossing the widom line, analogous to the subcritical phase change. Even, Banauti et al. [4] have suggested a modified phase diagram delineating the boundary based on the fluid's physical attributes in the supercritical region.

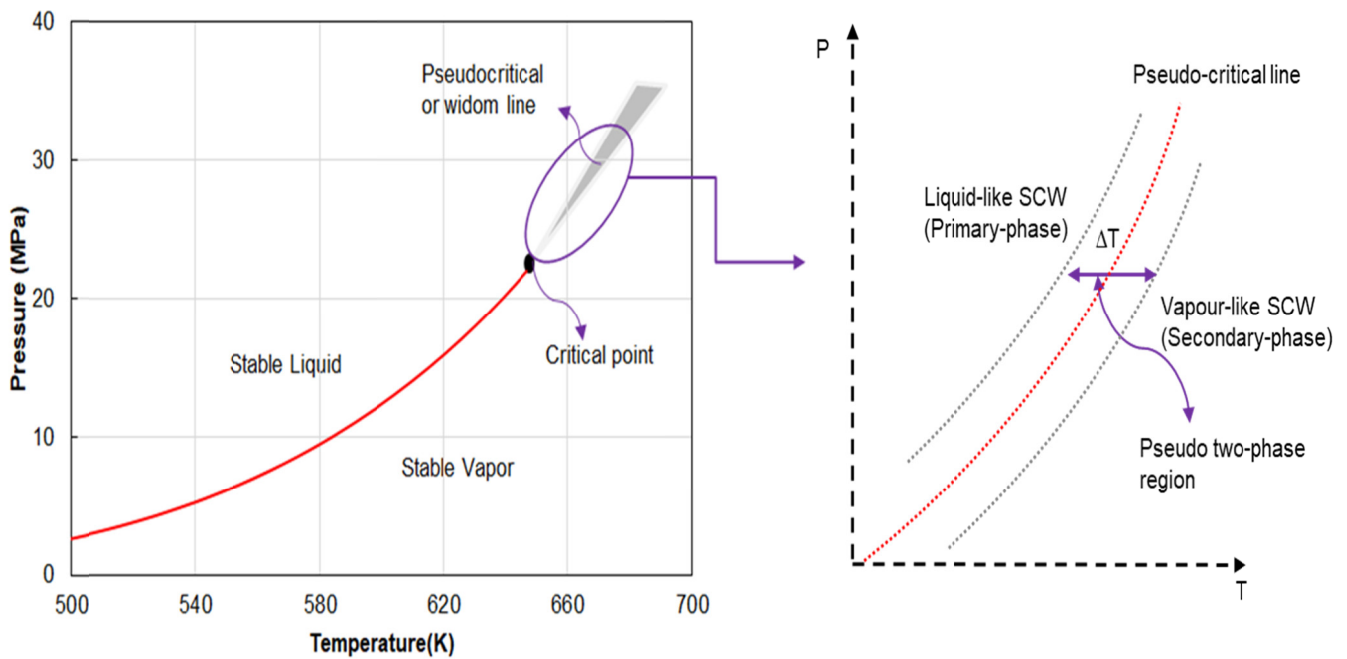

**Figure 1. Pressure Vs. Temperature for water including a schematic of the pseudo-two-phase region [1]**

Surprisingly, the subcritical phase-change correspondence to supercritical flow was conceived in the second half of the 20[th] century. Initial experimental studies [5–8] focused on heat transfer characteristics of supercritical fluid unanimously reported either heat transfer deterioration (HTD) or heat transfer enhancement (HTE). Shitsman [5] was the first to mention that the HTD occurrence might be caused by something similar to film-boiling. Meanwhile, Ackerman [7] coined the word pseudo boiling to underline the analogy of heat transfer behaviour of supercritical flow to normal boiling processes. However, later an alternate explanation based on the buoyancy forces originating out of the idiosyncratic variation of thermo-physical properties across the widom line came to light. Figure 2 depicts the properties of water with respect to temperature at pressure 25.3MPa.

Shitsman[9] reasoned the source of HTD might be due to a decline in transverse turbulent velocity fluctuation because of buoyant forces generated by radial density variation. Later, Hall and Jackson [10–12] asserted that the buoyancy force manipulates the velocity profile to reduce the turbulence production term near the wall. Nonetheless, some evidence contradicts the above explanations because the HTD was found to exist even in the absence of buoyancy forces at certain flow conditions. For instance, Shiralkar and Griffith [13]reported the decline in heat transfer for upward as well as downward flow. This paradox was resolved later when the origin of such HTD was ascribed to the acceleration effect instead of buoyancy. Numerical studies [14–19] have been performed for different flow conditions corresponding to pre-existing experimental setups and have shown consistency with the result by predicting HTD and HTE. Similar to experimental analysis, these studies have also found either buoyancy or acceleration effects to be the main reason for HTD occurrence.

Although, the very first explanation of the HTD [5][7] was conjectured based on the phase-change process where maxima are the result of the phenomena something similar to boiling in subcritical flows. But, the two-phase aspect of supercritical fluid flow remains unprobed. All the studies, either experimental or numerical, are still bound to the idea of treating flow as a single phase with varying property. Also, the current literature lags in terms of mathematical explanation for the plunge in the heat transfer characteristic of supercritical flow. Nonetheless, all the works have mentioned forces stemming from the peculiar property variation to be the reason for HTD. Still, no attempt has been made to quantify these variables in terms of forces. Similarly, no expression exists to estimate the region near the wall where most property variations are concentrated and give a vital length scale for the supercritical flow.

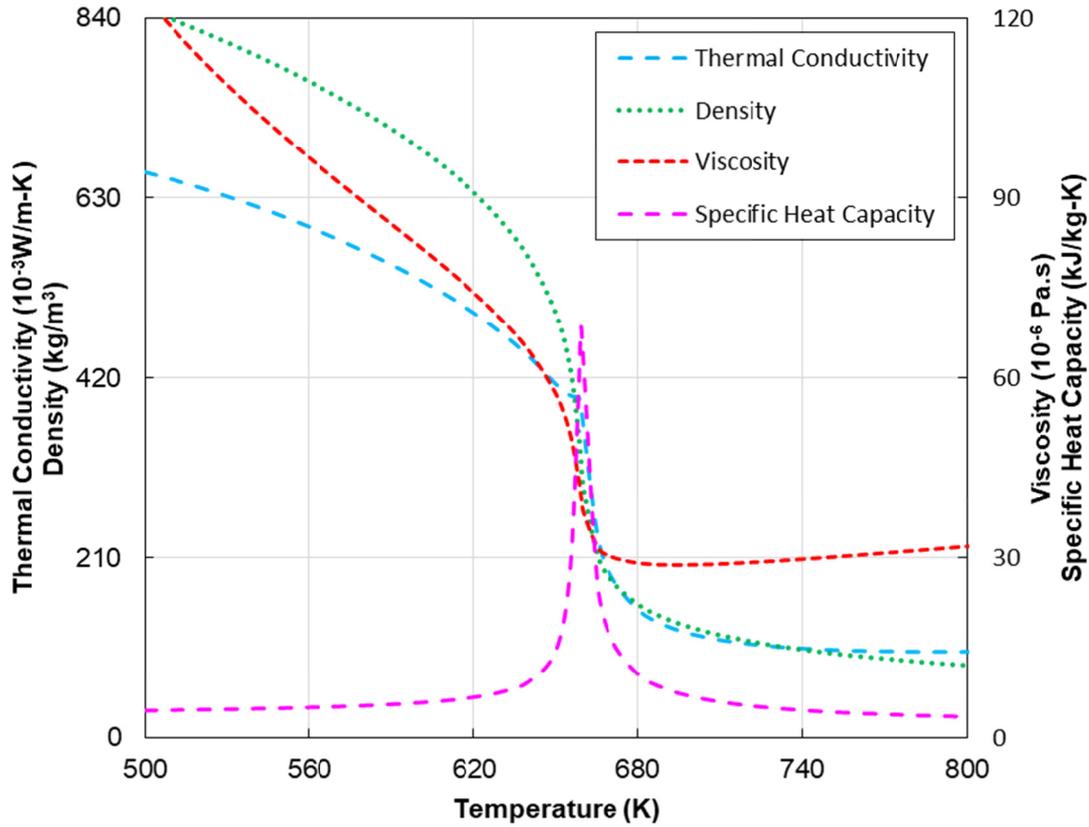

**Figure 2. Properties of water at 25.3MPa (NIST miniREFPROP)**

An attempt has been made to touch upon the lacunae, as mentioned earlier in the following way. A multiphase numerical model has been implemented where the flow is treated as a two-phase flow undergoing phase change across the pseudo-critical line, as shown in figure 1. Besides, a parametric study of the various thermos-physical properties has been performed to reveal the most dominant of all the properties on the flow behaviour alteration. Further, the flow is modelled for the quantitative analysis based on the parametric study's insight, and the mathematical analysis has explained the observed flow characteristics. In the end, volume fraction and dimensionless two-phase thickness (P) have been plotted for various flow conditions. The scaling analysis of h has been further extended to derive a theoretical expression for the same.

## 2. Methodology

## 2.1 Governing equations

The flow governing equation of mass, momentum, and energy is numerically solved for the traditional single-phase approach using commercial software (ANSYS FLUENT). The axis-symmetric flow for a steady-state case has been studied for the circular pipe flow. Accordingly, the governing equations have the radial and axial component, written in cylindrical coordinates as follows:

Continuity equation:

$$\frac{1}{r}\left\{\frac{\partial}{\partial x}(r\rho u) + \frac{\partial}{\partial r}(r\rho v)\right\} = 0 \tag{1}$$

Axial Momentum equation:

$$\frac{1}{r}\left\{\frac{\partial}{\partial x}(r\rho u^2) + \frac{\partial}{\partial r}(r\rho uv)\right\} \tag{2}$$

$$= -\frac{\partial p}{\partial x} - \rho g + \frac{1}{r}\frac{\partial}{\partial x}\left\{2r\mu_e\left(\frac{\partial u}{\partial x} - \frac{1}{3}(\nabla \cdot \vec{u})\right)\right\}$$

$$+ \frac{1}{r}\frac{\partial}{\partial r}\left\{r\mu_e\left(\frac{\partial u}{\partial r} + \frac{\partial v}{\partial x}\right)\right\}$$

Radial Momentum direction:

$$\frac{1}{r}\left\{\frac{\partial}{\partial x}(r\rho vu) + \frac{\partial}{\partial r}(r\rho v^2)\right\} \tag{3}$$

$$= -\frac{\partial p}{\partial r} + \frac{1}{r}\frac{\partial}{\partial r}\left\{2r\mu_e\left(\frac{\partial v}{\partial r} - \frac{1}{3}(\nabla \cdot \vec{u})\right)\right\}$$

$$+ \frac{1}{r}\frac{\partial}{\partial x}\left\{r\mu_e\left(\frac{\partial u}{\partial r} + \frac{\partial v}{\partial x}\right)\right\} - 2\mu\frac{v}{r^2} + \frac{2}{3}\frac{\mu}{r}(\nabla \cdot \vec{u})$$

$$\nabla \cdot \vec{u} = \frac{\partial u}{\partial x} + \frac{\partial v}{\partial r} + \frac{v}{r} \tag{4}$$

Here, $\mu_e$ is the effective viscosity defined as $\mu_e = \mu + \mu_t$, where $\mu_t$ is the turbulent viscosity determined by the turbulence model.

Energy equation:

$$\frac{1}{r}\left\{\frac{\partial}{\partial x}(r\rho uH) + \frac{\partial}{\partial r}(r\rho vH)\right\} \tag{5}$$

$$= \frac{1}{r}\left\{\frac{\partial}{\partial x}\left[rC_p\left(\frac{\mu}{Pr} + \frac{\mu_t}{Pr_t}\right)\frac{\partial T}{\partial x}\right] + \frac{\partial}{\partial r}\left[rC_p\left(\frac{\mu}{Pr} + \frac{\mu_t}{Pr_t}\right)\frac{\partial T}{\partial r}\right]\right\}$$

$Pr_t$ is taken to be equal to 0.85 for this analysis. The SST k-ω turbulence model is used for this study, which has the advantage that it subsumes the qualities of the k-ω model and the k-ε. It has the properties of k-ω in wall proximity; meanwhile, it behaves like k-ε in the free stream region.

The volume of fluid (VOF) multiphase model has been used for two-phase simulation. The above model is generally used to track the interfaces between immiscible phases by calculating all the phases' volume fractions in each computational cell. But the recent developments [20–22] have proved its suitability for the flow boiling process, which involves mass transfer across the phases. So, the only modification in the governing equation is that the continuity equation is solved individually for the volume fraction of the secondary phase, which is vapor in our case. Hence, the equation takes the following form:

$$\frac{1}{\rho_v}\left[\frac{\partial(\beta_v\rho_v)}{\partial t} + \nabla\cdot(\beta_v\rho_v\vec{u}_v) = (\dot{m}_{lv} - \dot{m}_{vl})\right] \tag{6}$$

Here, $\dot{m}_{vl}$ or $\dot{m}_{lv}$ is the mass transfer across the phase due to condensation or evaporation. These rates are governed by the Lee model whose detailed explanation can be found in Ansys fluent theory guide. Since the continuity equation for the secondary phase is only solved, and the volume fraction for the primary phase (liquid) is calculated subject to the constraint shown below:

$$\beta_l + \beta_v = 1 \tag{7}$$

All the other governing equations such as momentum and energy are solved for the fluid as a whole (same as single fluid equations which are given earlier) where all the properties are taken as volume averaged, having a contribution from every phase present in the computational cell. In other words, there is a single momentum as well as an energy equation for the multiphase flow with modified properties accounting for the presence of different phases. Consequently, all the flow variables computed from the governing equation incorporates the existence of all the phases. For example, enthalpy of evaporation or condensation has been accounted in such a way that the enthalpy difference between the vapor and liquid phase at saturation temperature is equal to the latent heat of phase change.

$$H_v^{sat} - H_l^{sat} = \text{(Latent heat of phase change)} \tag{8}$$

## 2.2 Physical and Numerical model

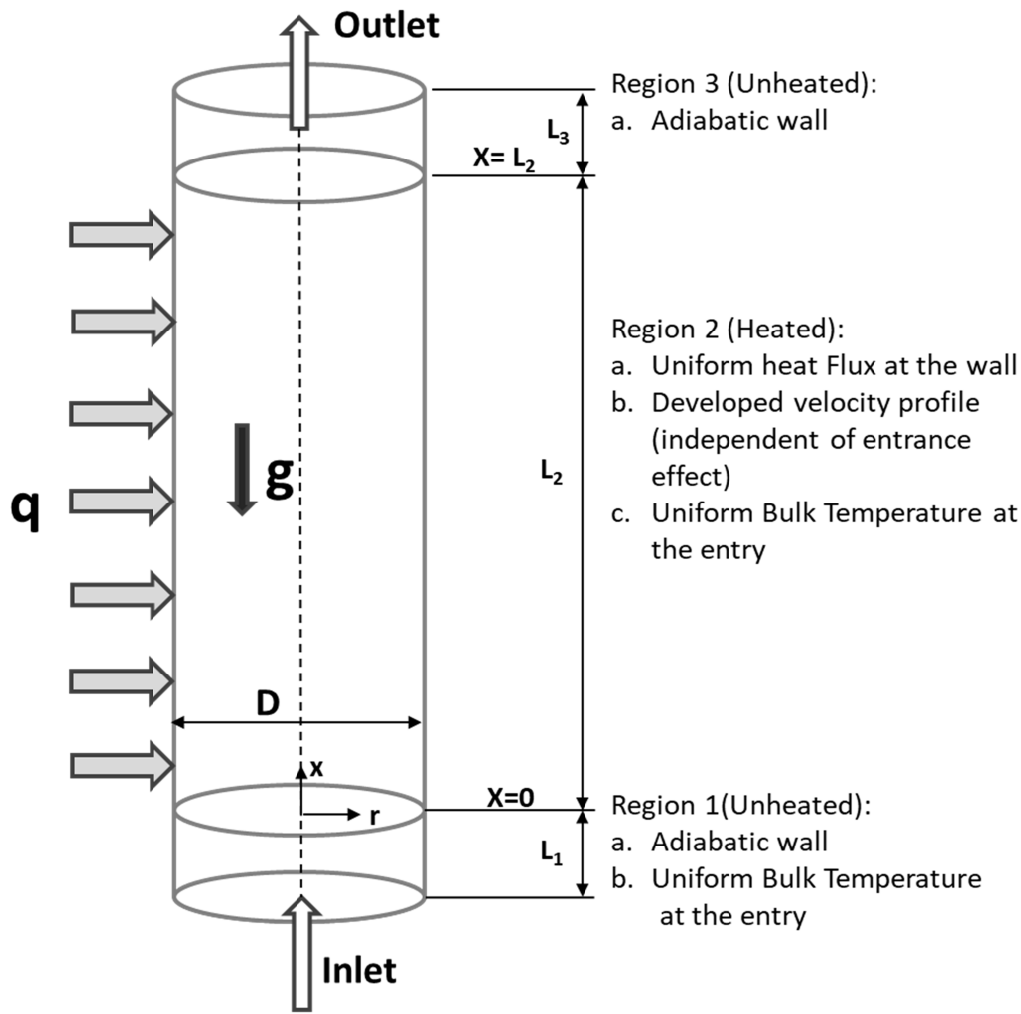

**Figure 3. Schematic diagram of flow geometry**

Figure 3 shows the general representation of the computational domain where the circular pipe's length and diameter vary depending on the experiments, as shown in Table 1. The flow geometry used is a vertical circular pipe, and for the numerical analysis purpose, it is partitioned into three different regions along the axis. The first part is region 1, where no heat transfer occurs and results in developing a velocity profile overcoming the entrance effects. Next is region 2, which imitates the test section's heated length (uniform heat flux) in various experimental setups. At the end, region 3 with an adiabatic wall ensures no backward flow, leading to a stable numerical system.

The inflated radial gradient of every variable exists near the boundary as a no-slip condition is maintained at the flow domain wall. All the pre-existing works suggest that the $y^+$ of the first element near the wall should be less than 1. This has been accommodated by introducing radial non-uniformity in the mesh such that the highest grid density exists in the wall neighbourhood. The distance of the first grid from the wall is kept at 0.001 mm to maintain the value of $y^+$ less than 0.2, so that it falls within the recommended limit.

The property variation across the pseudo-critical line resembles the phase-change process in the subcritical fluid, as shown in figure1. So, to set an analogy with the conventional boiling process, we have assumed a virtual phase-change process across the pseudocritical line and approached the same using the VOF model. The phase transition is implemented by the mass transfer mechanism dictated by the Lee model. It is a condensation-evaporation model with a user defined coefficient (or frequency), and for the current study, the value of 2500 and 10 are chosen for evaporation and condensation, respectively. The latent heat of phase change (enthalpy of evaporation or condensation) is decided by the $\Delta T$ shown in figure 1 and the average specific heat over the $\Delta T$ interval.

$$\text{Enthalpy of evaporation(or condensation)} = \text{Avg.}(C_p) \times \Delta T \qquad (9)$$

For the present study, $\Delta T$'s value is chosen to be 2 with equally distributed on both sides of psuedocritical temperature since it is set as the saturation temperature. But it is essential to point out that there is no method of choosing $\Delta T$, which introduces a default parameter in the model that has to be decided.

National Institute of Standards and Technology (NIST) software miniREFPROP has been referred for water's thermophysical properties at various operating conditions. In the simulation, a piecewise linear profile is used to incorporate the properties as a function of temperature, assuming constant pressure. Since the pressure drop is significantly less, so it is a valid assumption. The geometric details for both the case are given in Table 1.

**Table 1. Geometry specification for the computational domain**

| Geometry | Cross-section (D) | Unheated Length ($L_1$) | Heated length ($L_2$) | Unheated length ($L_3$) | Total length |
|---|---|---|---|---|---|
| Circular (Shitsman) | 8 mm | 100 mm | 1500 mm | 100 mm | 1700 mm |
| Circular (Ornatskij) | 3 mm | 100 mm | 800 mm | 100 mm | 1000 mm |

## 2.3 Boundary and Operating condition

The boundary and operating conditions are given in Table 2 for all the cases. The flow conditions are taken from Shitsman's [5] and Ornatskij [19] experiments, where HTD has been imputed to buoyancy and acceleration effects, respectively. At the inlet boundary, turbulent intensity and turbulent viscosity ratio are set as 5% and 10 each. Since the fluid is compressible, mass flow inlet and pressure outlet have been applied as the boundary conditions. The gravitational acceleration (g) value is taken to be $9.8 m/s^2$.

**Table 2 Boundary and Operating condition**

| Parameter | Case-1 | Case-2 | Case-3 |
|---|---|---|---|
| Reference | Shitsman's experimental setup with HTD having two peaks in wall temperature | Ornatskij (1971) experimental setup with HTD | Shitsman's experimental setup with no HTD |
| $T_{pc}$ | 659 K | 659 K | 652 K |
| G | 0.022 kg/s | 0.0106 kg/s | 0.022 kg/s |
| $T_{in}$ | 578 K | 500 K | 600 K |
| q | 384.8 kW/m$^2$ | 1810 kW/m$^2$ | 220.8 kW/m$^2$ |
| $P_{out}$ | 25.3 MPa | 25.3 MPa | 23.3 MPa |

# 3. Results and Discussion

## 3.1 Grid independence

All simulation results have been ensured to be grid independent. It has been achieved by refining the mesh such that node points are doubled for every succeeding study. Mesh has been refined in radial as well as in the axial direction. Initially, the number of nodes was kept at 200000 and 60000 for case-1 and case-2, respectively, and then the grid was modified subsequently. Although mesh is refined radially, the distance of the first grid point near the wall is kept the same in all the simulations to maintain the desired $y^+$ value.

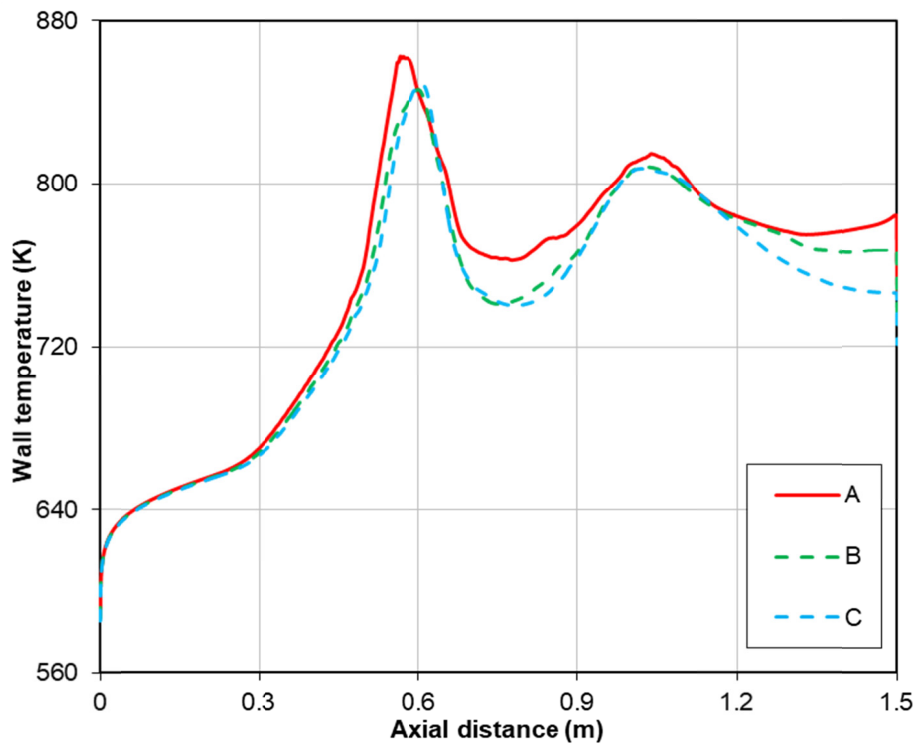

(a) A=200000 nodes, B=2*A and C=2*B

Figure 4 (a) and (b) show the wall temperature variation for case-1 and case-2 during the grid independence study. The plot reveals that the result achieves grid independence at B number of nodes for both the cases. Also, it has been noticed that the wall temperature is insensitive to the grid refinement in the axial direction. Based on the above observation, the final mesh structure was decided to balance accuracy and computational cost. So, all the simulations are done on the grid structure with the specification are dx=0.6mm and dr=0.03mm for both the cases, leading to 447000 nodes and 123000 nodes for case-1 case-2, respectively.

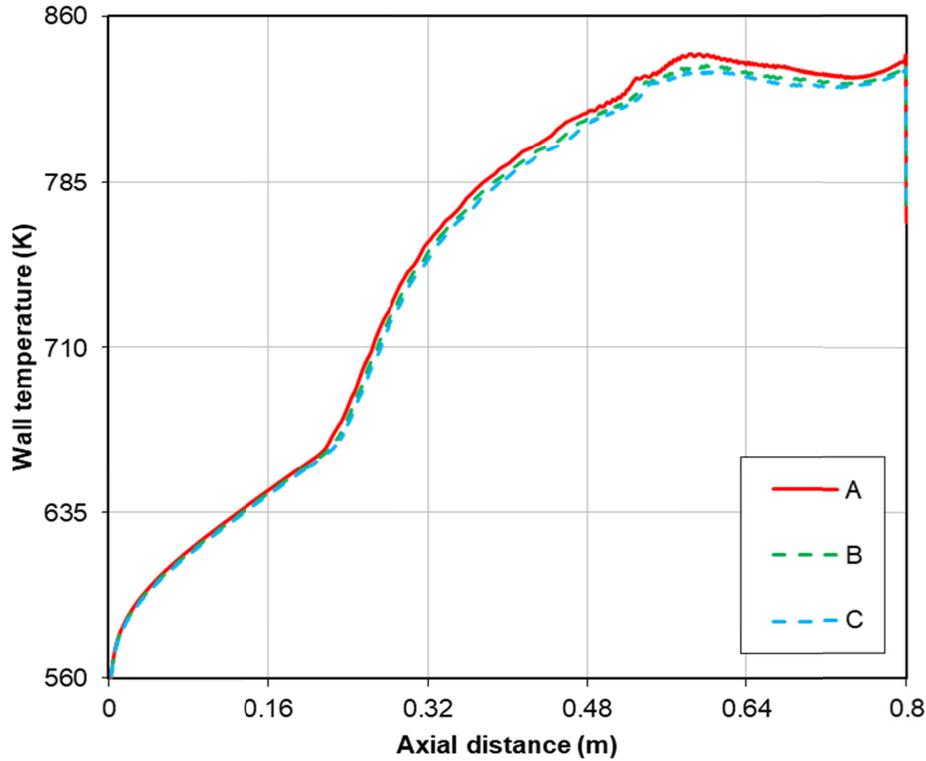

**(b)** A=60000 nodes, B=2*A and C=2*B

**Figure 4. Grid independent plot for different cases (a) Case-1 (b) Case-2**

## 3.2 Comparison with the conventional method

The numerical analysis has been performed for the proposed model (two-phase) and the conventional approach (single-phase) to show the two methods' compatibility. Also, to establish a comprehensive application of the propounded model, case-1 and case-2 have been studied. Although both cases correspond to the HTD phenomena, the source responsible for the impairment of heat transfer is very different. Figure 5 exhibits a comparison of numerical and experimental results for both cases. The numerical analysis is further bifurcated in the two approaches mentioned above. The juxtaposition reveals that numerical analysis is competent enough to predict the right trend. However, there are slight inconsistencies in the magnitude, and the location of the HTD symbolised by the wall temperature peaks.

It is a well-established fact that any turbulence loss directly results in downgrading the convective heat transfer, which largely depends on fluid mixing. So, for HTD to happen,

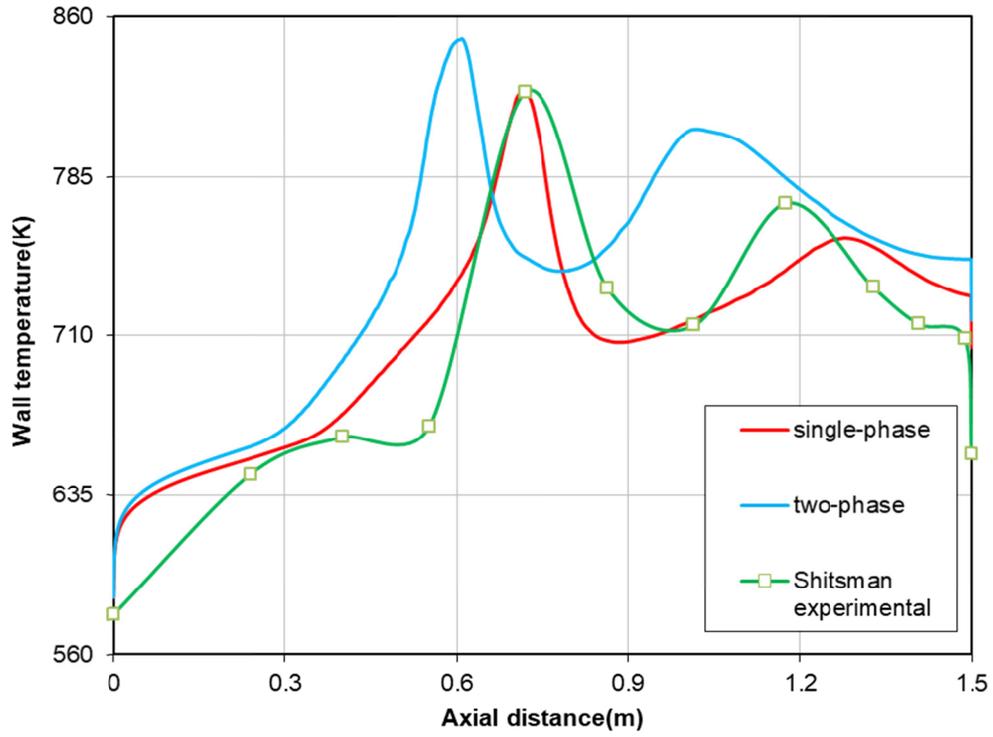

(a)

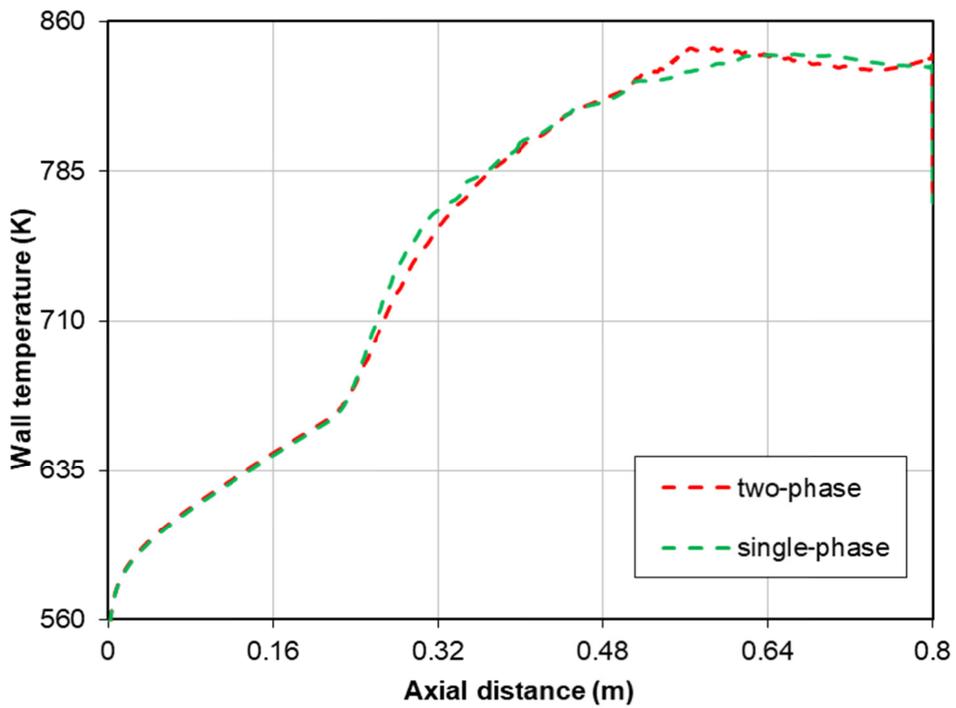

(b)

**Figure 5. Wall temperature variation (a) Case-1 (b) Case-2**

.

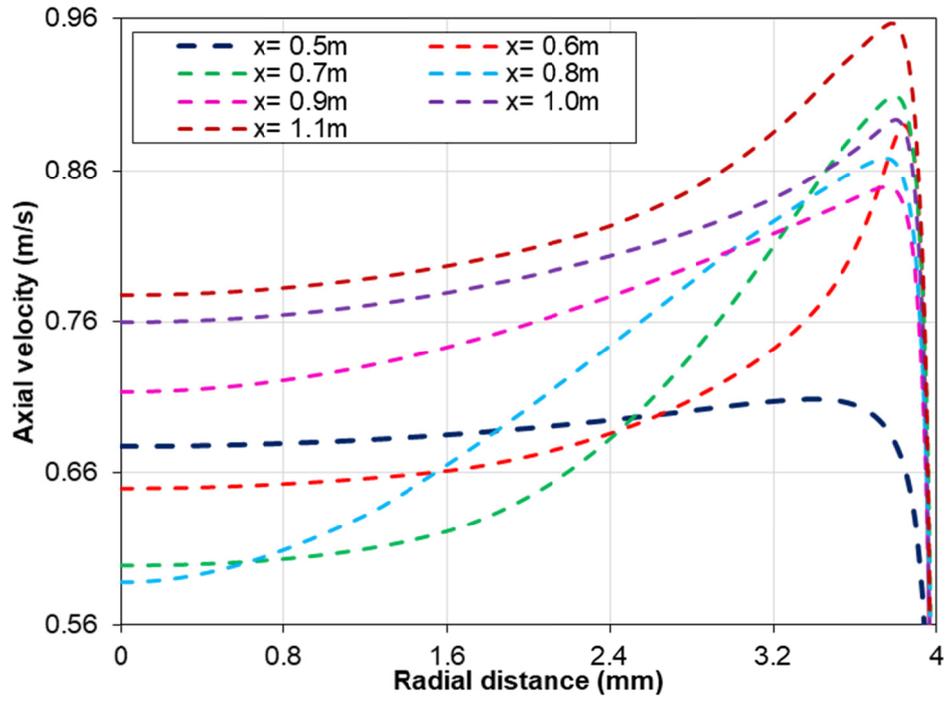

**(a)**

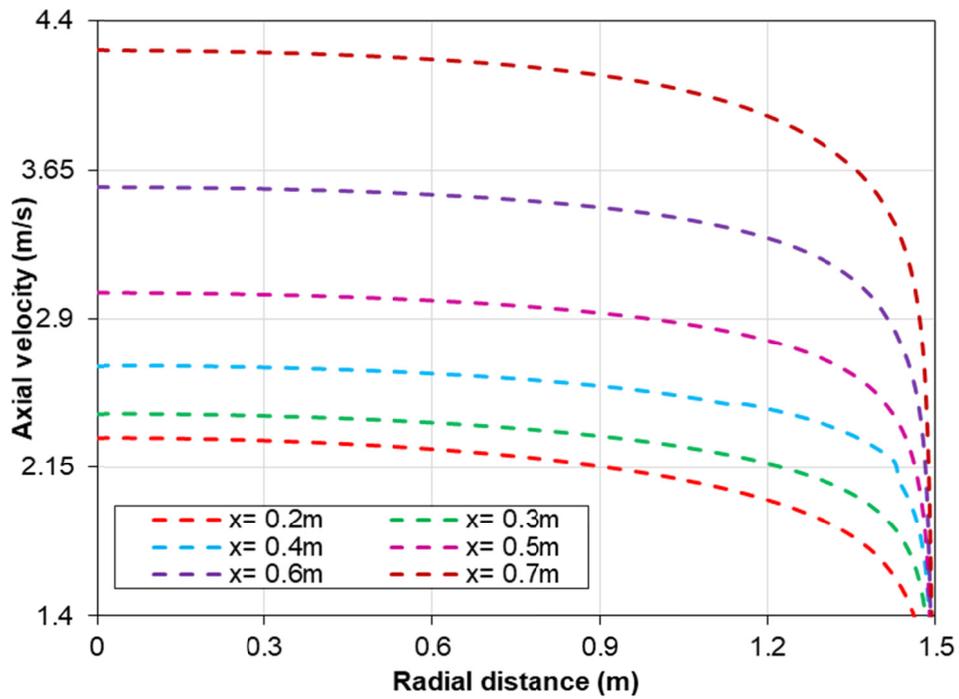

**(b)**

**Figure 6. Velocity profile at different axial location (a) Case-1 (b) Case-2**

there must be a plunge in the TKE, which is the best judging parameter for the flow's turbulence level. Also, the forces originating out of the fluid property disparity alter the flow, which is reflected in the velocity plots and since, TKE production term is expressed in terms of rate of strain tensor modus. Therefore, the slope of the radial velocity profile qualifies to be the best analysis parameter. All this leads to the conclusion that any depletion of the velocity gradient will be reflected as a decline in the TKE value.

Figure 5(a) corresponds to the case-1, which is Shitsman's experimental setup, where two wall temperature peaks were observed, and peculiarly HTD reoccurred after showing signs of recovery. The numerical study matches the observed trend, although the two-phase and single-phase approach does not imitate each other perfectly. However, this can be anticipated because the VOF model and the pseudo-phase change process has a lot of parameters that need to be properly fine-tuned. In the neighbourhood of the first peak, such as x=0.5m velocity profile is flattened near the wall, as shown in figure 6(a) because of buoyancy forces arising out of the radial density variation. Meanwhile, as expected TKE values experience a drop near the first peak, as displayed in figure 7(a). As the heat transfer recovery starts from the first peak location, the velocity profile suggests that buoyancy forces are still dominant in the near wall region, eventually leading to higher velocity near the wall. This is also translated in the TKE plot, where the value keeps on increasing near the wall.

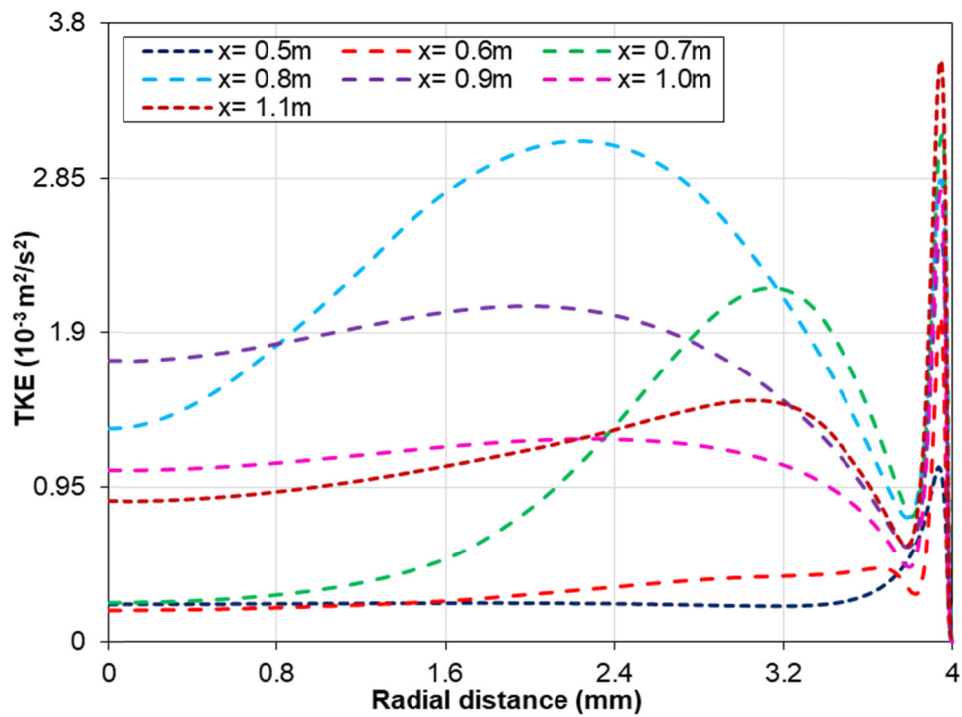

(a)

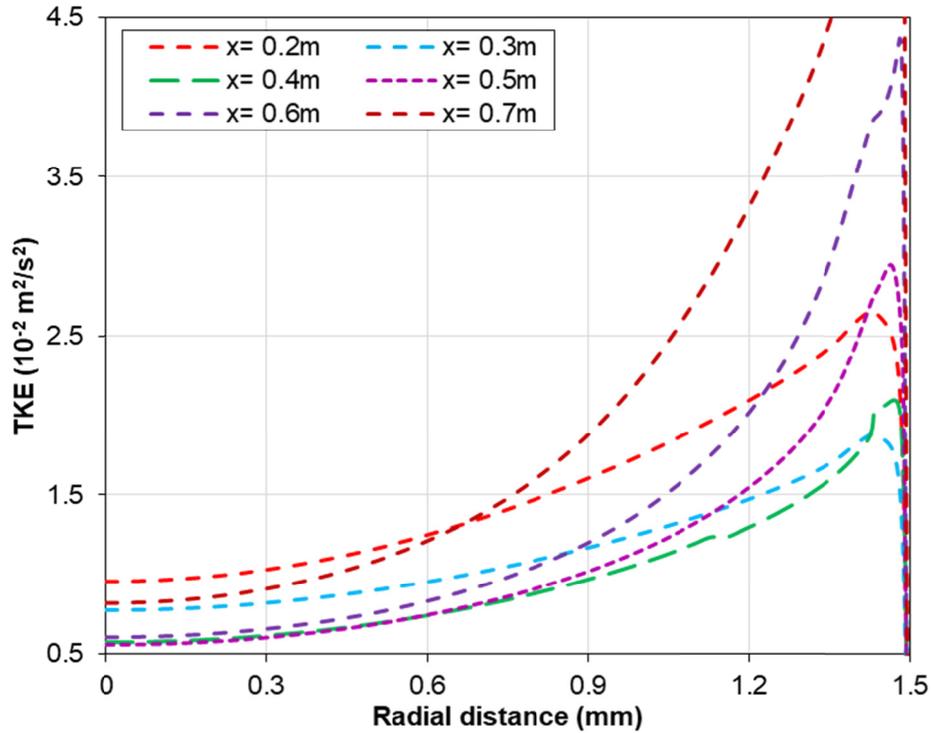

(b)

**Figure 7. TKE profile at the different axial location (a) Case-1 (b) Case-2**

Further downstream, the flow experiences an increase in the bulk velocity, such as for x=0.9m, and gradually this results in flattening of the velocity profile away from the wall in the proximity of the second wall temperature peak. It leads to loss of TKE majorly in the bulk region at x=0.9m and x=1.0m, which eventually results in HTD. The reoccurrence of the heat transfer impairment is primarily due to the acceleration effects where the bulk fluid rushes to maintain the same mass flow rate despite the decreasing density value in the axial direction.

In addition, case-2, inspired by the Ornatshij experiment, has been studied, and figure 5(b) represents the two numerical approaches that produce very similar results. Contrary to case-1, mass flux and heat flux are much higher in this analysis; thus, flow behaviour is much different even though the fluid has the same property. This study falls into the category of HTD, which is solely the consequence of the acceleration effect. The rapid decrease in the fluid density propels the flow, as displayed in figure 6(b), where the velocity keeps on increasing downstream. As a result, the velocity gradient decreases in bulk, marked by the decline in TKE value in the bulk even though it grows near the wall, as manifested in figure 7(b). So, a wall temperature peak is observed, which is figurative of HTD.

In the end, the wall temperature maxima characteristic correlates with the nature of the forces responsible for the HTD. The two extremes of the temperature peaks are marked as the sharp one resulting from buoyancy forces and broad ones which are the consequence of the acceleration effects. So, all in all, we can say that the introduced two-phase method is in agreement with the conventional single-phase understanding as it unveils the HTD in a similar fashion.

## 3.3 Parametric study of properties

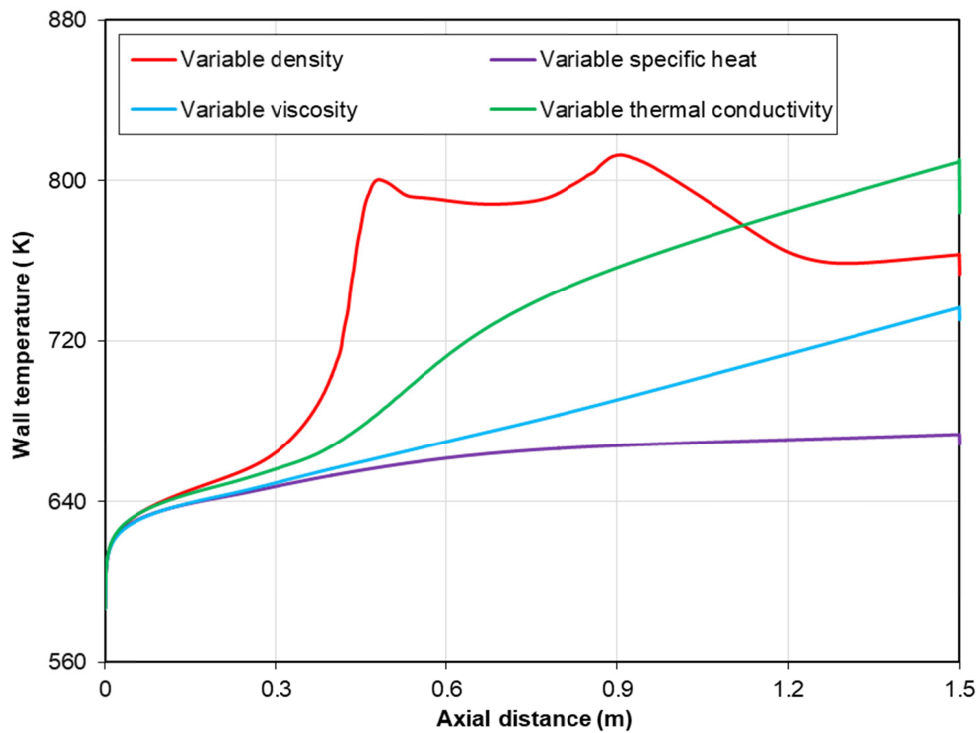

(a)

The parametric study's main objective is to unearth the most dominant one of the four fluid properties dictating the flow behaviour. For this purpose, four simulations corresponding to every property have been performed using the conventional numerical setup (single-phase), retaining everything unchanged except the properties. For each run, one of the properties is kept variable similar to the original flow; meanwhile, the rest are maintained constant at their inlet value. In other words, property variation becomes the parameters in this parametric study.

Figure 8(a) corresponds to case-1, which represents that the density variation has the most significant influence on the flow behaviour followed by the thermal conductivity, whereas specific heat has the least impact. Interestingly, the variable density simulation predicts two wall temperature peaks that are there for the actual flow.

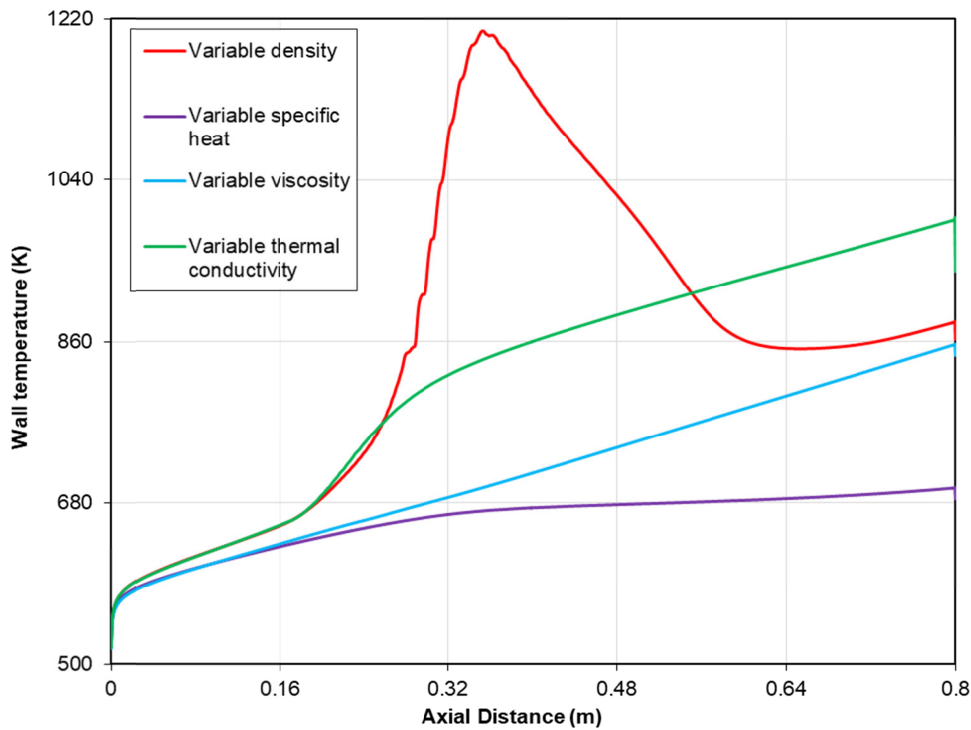

**(b)**

**Figure 8. Result of parametric study (a) Case-1 (b) Case-2**

Similarly, figure 8(b) represents that for case-2 density difference, only can forecast the HTD so establishing the dominance of density over the flow behaviour like the first case. Also, the properties governing the flow characteristic follows the same order as observed in case-1. This insight into the supercritical flow is somewhat expected from our qualitative understanding because all the forces altering the flow field, such as buoyancy or the acceleration effects, are basically the direct consequence of the density disparity. If we extrapolate this, we can observe supercritical flow behaviour even in the subcritical flow, where density varies over a wide range.

It is worth mentioning that if we use the different values of the constant properties instead of their inlet value, we will get a totally different result where the HTD may not be as prominent as it is for this particular study. But even all those possible analysis will suggest that the density has the highest control over the flow behaviour, and the flow will be closest to the actual flow for the variable density setup.

## 3.4 Quantitative analysis of flow

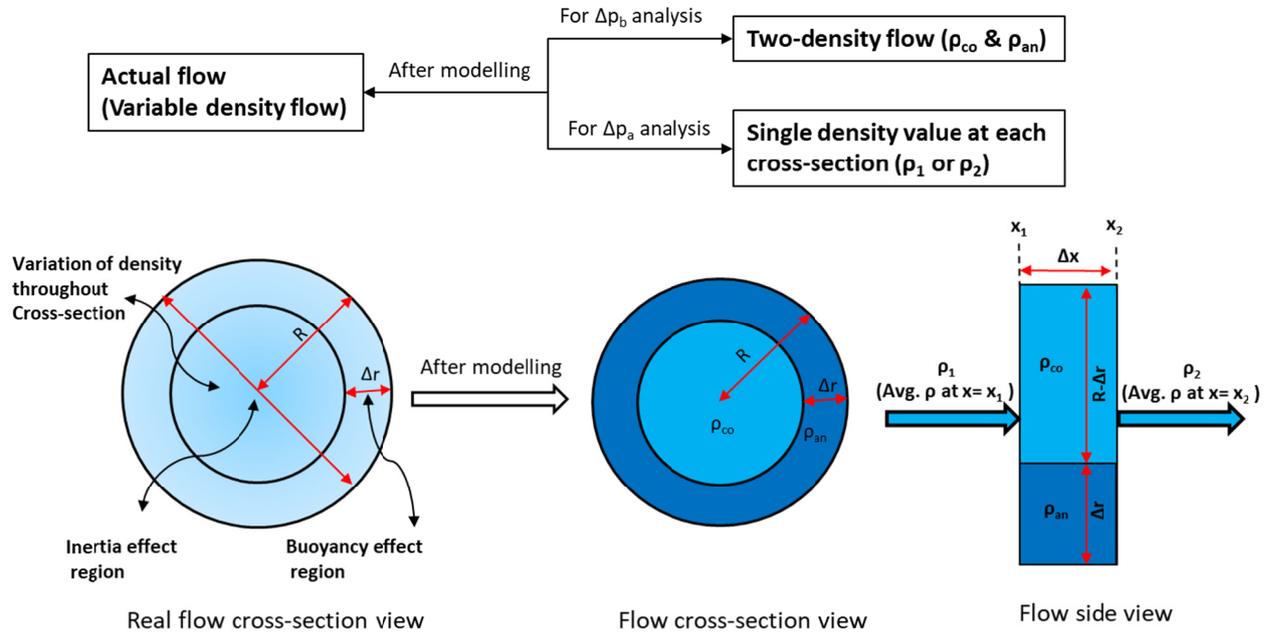

**Figure 9. Schematic of flow simplification for analysis**

The quantitative analysis aims to gauge the different forces numerally originating from the density difference in the axial and radial direction. As mentioned in the preceding sections, buoyancy and acceleration effects are the two broad classifications of forces responsible for the heat transfer impairment in supercritical flows. So, the actual flow has been simplified to calculate the magnitude of the two forces on a small axial element of $\Delta x=0.01$m as follows:

1. The major share of the buoyancy force is experienced in the near wall region, where the flow has the highest radial density gradient. As displayed in figure 9, $\Delta r$ is the cross-sectional area where the buoyancy forces are predominant. Based on the following equations, $\Delta r$ is estimated for this study.

$$(\rho_{r-\Delta r} - \rho_w) = 0.9 * (\rho_c - \rho_w) \quad (10)$$

Where $\rho_c$ = density at the center, $\rho_w$ = density at the wall, and $\rho_{r-\Delta r}$ = density at a radial distance of r-$\Delta r$ from the center. Therefore, employing the above definition of $\Delta r$, the flow has been modelled as if it consists of two density regimes in the cross-section defined as following:

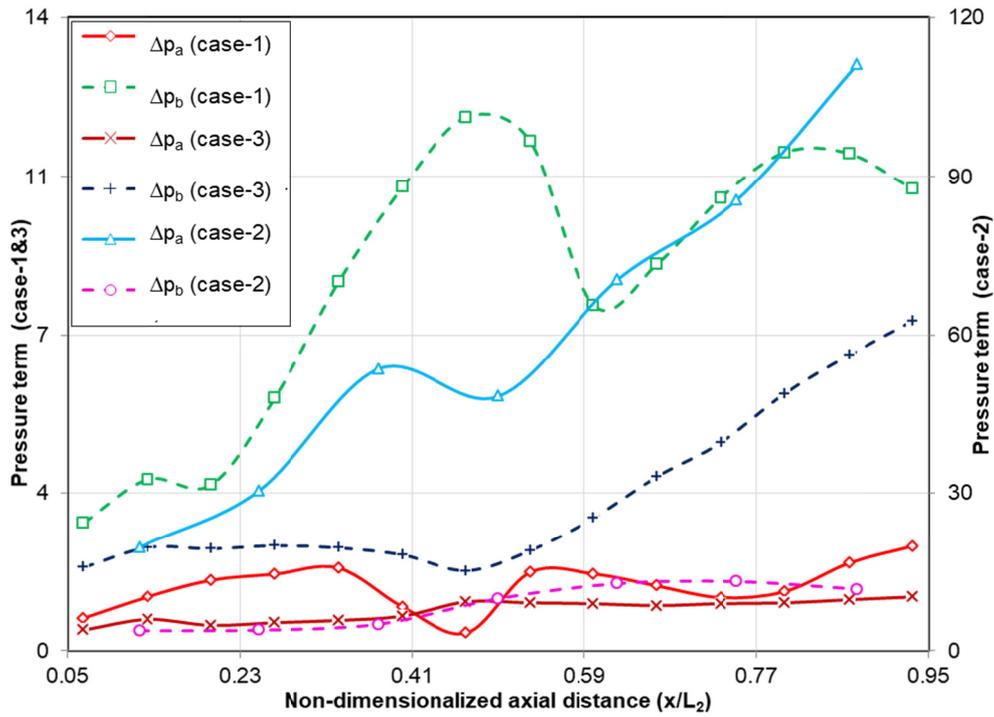

(a)

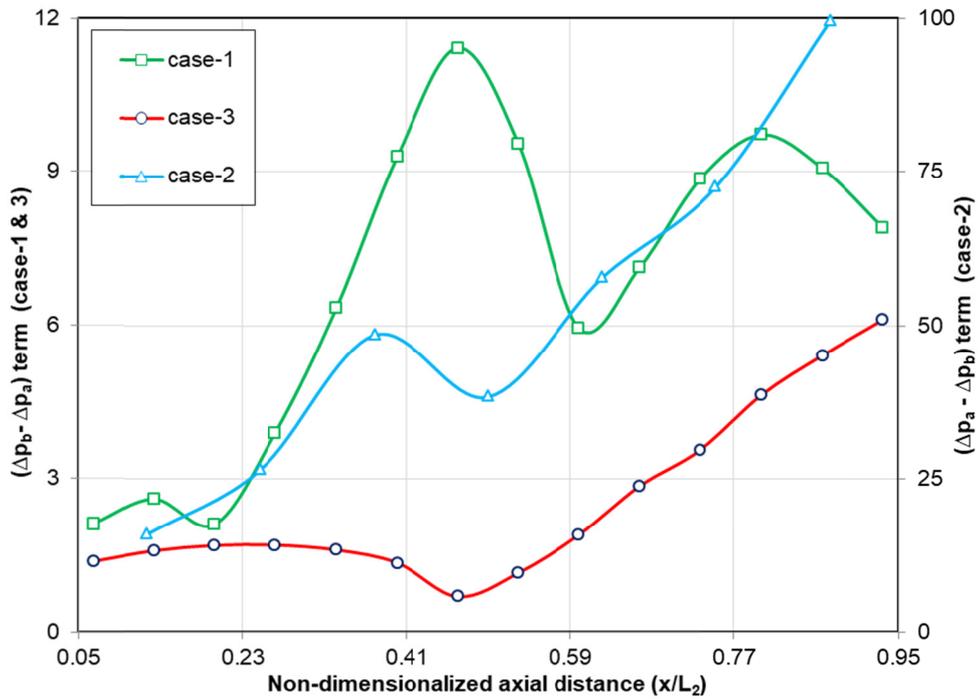

(b)

**Figure 10. Pressure terms axial plot (a) Pressure terms variation for all cases (b) Difference between the two pressure components depending on the dominant term**

$$\rho_{c0}(\text{core region}) = \text{Avg.}(\rho_c \text{ to } \rho_{r-\Delta r}) \qquad (11)$$

$$\rho_{an}(\text{annular region}) = \text{Avg.}\,(\rho_{r-\Delta r} \text{ to } \rho_w) \qquad (12)$$

Accordingly, net buoyancy forces are defined in terms of force per unit cross-sectional area (CA) using equation (13) with an assumption that radial density profile remains the same throughout the small axial element length ($\Delta x$) considered for analysis.

$$\Delta p_b = \frac{F_b}{CA} \sim (\rho_{co} - \rho_{an})*g*\Delta x * \frac{CA_{\Delta r}}{CA} \qquad (13)$$

2. The fluid experiences acceleration effects throughout the cross-section because its genesis lies in the axial density variation. In other words, the more is the density gradient downstream; the higher is the fluid acceleration to maintain the same mass flow rate. As displayed in Figure 9 (flow side view), actual flow has been restructured to have a single density value at every axial location, which is the average cross-sectional density. Hence, acceleration effects are enumerated as the difference in the inertia force per unit cross-sectional area shown in equation (14).

$$\Delta p_a = \frac{F_a}{CA} \sim (\rho_{x_2} U_{x_2}^2 - \rho_{x_1} U_{x_1}^2) \qquad (14)$$

So, after some manipulation, equation (14) takes the following form

$$\Delta p_a \sim G^2 \left( \frac{\rho_{x_1} - \rho_{x_2}}{\rho_{x_1} * \rho_{x_2}} \right) \qquad (15)$$

Here $U_{x_1}$ = average axial velocity at $x = x_1$ and $U_{x_2}$ = average axial velocity at $x = x_2$
Further, the above expressions of the forces have been utilized to calculate the value of the forces at various axial locations along the flow path for all the cases. This probe has also been made more inclusive by performing the computation for case-3, which corresponds to Shitsman's [2] experiments, which reported no HTD. Figure 10(a) and 10(b) presents pressure term and the difference in the pressure terms, respectively, for all the cases. It elucidates our current understanding that even though both forces are present in the flow, the dominant one governs the flow. For example, in case-2 $\Delta p_a$ is at least one order higher than the $\Delta p_b$ throughout the flow, whereas for case-1, such clarity does not exist. In other words, at the location of the first HTD in case-1 $\Delta p_b$ exceeds the $\Delta p_a$ comfortably but for the second HTD as discussed earlier later has a greater influence than the former. Since at the spot of the second wall temperature peak $\Delta p_b$ passes through a local minimum; meanwhile the $\Delta p_a$ experiences a local maximum which is also reflected as a local nadir point in the pressure difference plot, as shown in figure 10(b). Surprisingly, case-3 has both the forces of more or less of the same magnitude in the entirety of the flow domain. This leads us to a significant deduction that the interplay of both the forces dictates the supercritical flow behaviour. That is to say that buoyancy and acceleration affect dominance marks the two extremes of the supercritical flow spectrum with HTD phenomena. In contrast, when the two forces are commensurate, no HTD is reported. The explanation lies in the way the velocity profile is tempered in the presence of the two forces, as shown in figure 11. Since the buoyancy is concentrated in the near wall region while the bulk experiences the major chunk of acceleration effect, their consequence is distributed over the cross-section, and the ideal behaviour lies in the balance of the two factors. The heat transfer coefficient plot in figure 11 has used the following equations.

$$T_b = \frac{\int_0^R 2\pi T C_p \rho u r \, dr}{\int_0^R 2\pi C_p \rho u r \, dr} \tag{16}$$

$$\text{heat transfer coefficient} = \frac{q}{T_w - T_b} \tag{17}$$

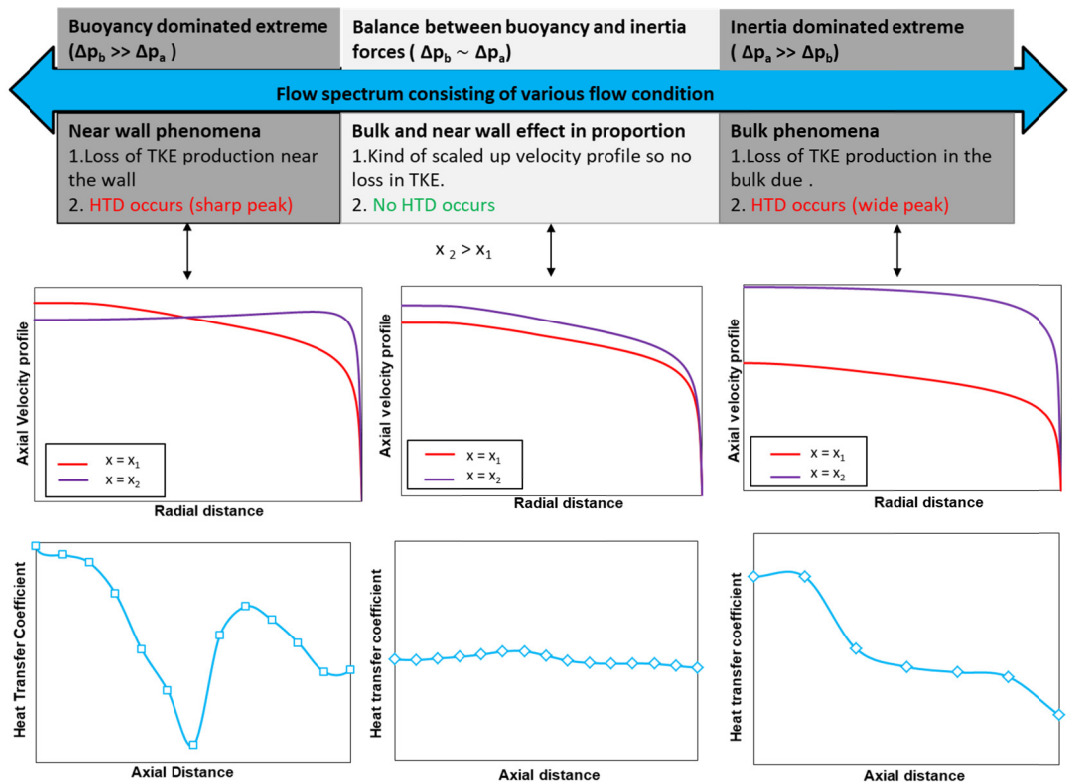

**Figure 11. Schematic depicting the various possibility of flow in supercritical fluids**

## 3.5 Volume fraction and interpretation

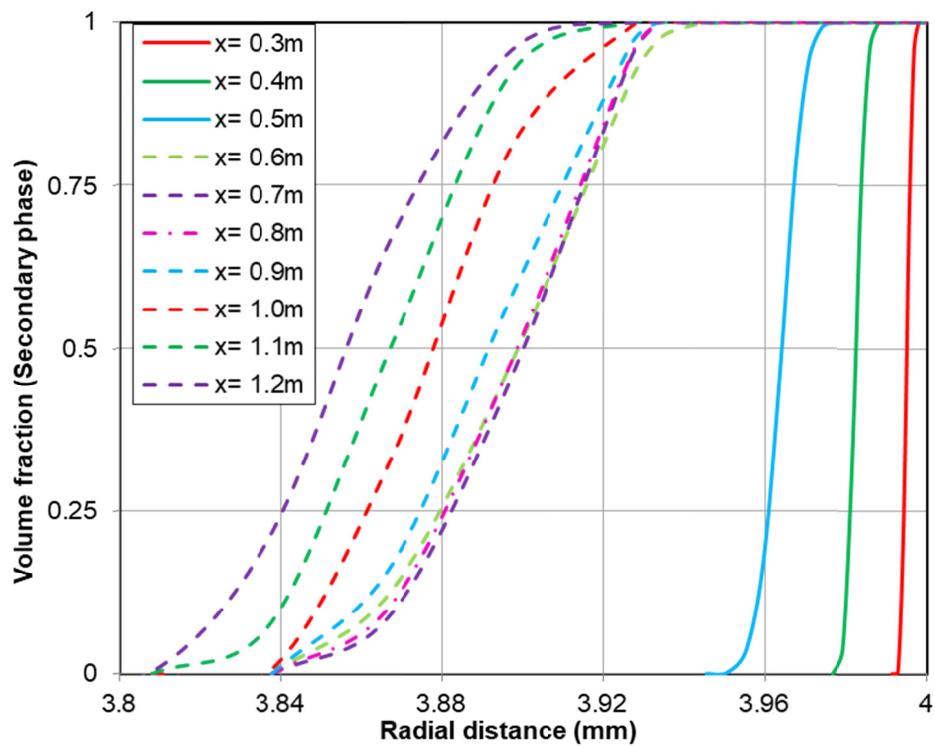

(a)

The most prominent edge of the pseudo phase change model over the single-fluid approach is that it provides us a new variable called volume fraction that can unveil the heat transfer characteristic to a greater depth. In other words, a two-phase system gives a better insight into the flow composition apart from the regular flow structure. It equips us with the phase distribution in the flow, which can be used to explore the analogy of supercritical flow with the boiling process in subcritical fluids.

Further, it is necessary to see the variation of volume fraction in all the cases before reaching any conclusion because the given phase distribution may be a normal phenomenon in supercritical flow. The volume fraction of the second phase (lighter fluid) for case-2 and case-1 has been mapped in Figures 12(a) and 12(b), respectively. The variations are plotted close to the wall because the volume fraction vanishes to zero in a short distance from the heated boundary. Case-2 shows a gradual variation of second phase penetration in the bulk fluid, which is similar to flow boiling where vapour height near the wall along the axial length increases gently. On the contrary, the volume fraction of the lighter fluid shows a sudden jump in penetration for case-1 at the location of the first peak. Figure 12 (a) displays this behaviour where at x= 0.6 m, the plot shows a steep rise in the value compared to volume fraction variation at x = 0.5 m. No such behaviour is observed for the second HTD location, and the difference lies in the nature of the forces responsible for its occurrence. These observations can be seen in analogy with the film boiling phenomena of subcritical fluids. In other words, the moderate increase in volume fraction along the length can be seen as somewhat similar to normal boiling (such as nucleate boiling) as in case-2. So, the transition from normal boiling to film boiling can be marked by the steep increase in volume fraction reflected at the first temperature peak position of case-1. So, one can use the term of pseudo film boiling phenomenon for the HTD caused by buoyancy forces solely but not for the one reported due to acceleration effects.

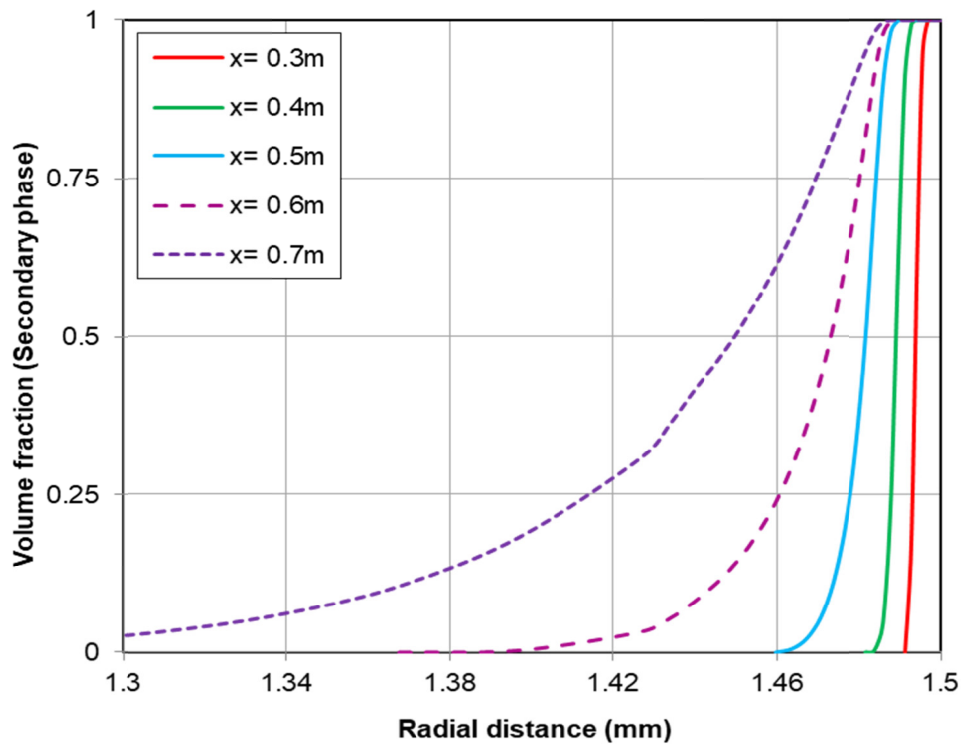

(b)

**Figure 12. Radial variation of volume fraction (Secondary phase) at different axial location (a) Case-1 (b) Case-2**

In addition, a scaling analysis has been done for the theoretical calculation of two-phase thickness (h), as shown in figure 13. This development aims to accommodate all the factors affecting the pseudo-phase change process and, eventually, present h into a single dimensionless expression. It has been approached in the following way:

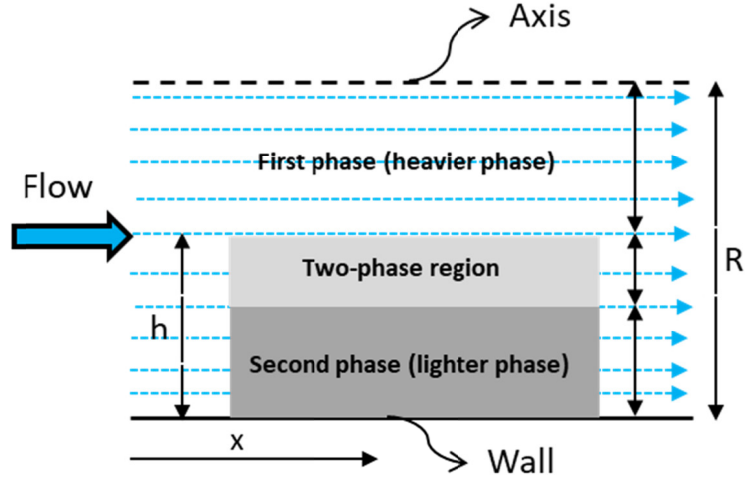

**Figure 13. Flow representation near the wall**

$$h \propto \begin{cases} \text{heat input } (q * SA) \\ 1/\text{mass flow rate } (m * CA) \\ 1/\text{Inlet enthalpy difference } (\Delta H = H_{pc} - H_{in}) \\ \text{radius } (R) \\ \Delta\rho/\rho_{pc} = (\rho_{in} - \rho_{ps})/\rho_{ps} \end{cases} \quad (18)$$

In the above equation, $\Delta H$ has been included since $T_{ps}$ has been set as the saturation temperature. Also, the last factor presence incorporates the dynamic nature of flow on the pseudo-phase change because $\Delta\rho$ term (assumed to be of the order $\rho_{in} - \rho_{ps}$) is common in the expression (13 and 15) of both the forces governing the supercritical flow. So, simplifying equation (18) keeping in mind that h is a function of x, therefore using $S.A = \pi * 2R * x$, $C.A = \pi * R^2$ and grouping all the variables involved in a nondimensional form

$$\frac{h}{R} = C * \left(\frac{qx}{mR\Delta H}\right) * \left(\frac{\Delta\rho}{\rho_{pc}}\right) \quad (19)$$

Here, C is a proportionality constant. After rearranging the above equation, it expresses a dimensionless number (P) representing the two-phase thickness.

$$P = \left(\frac{h}{R}\right)\left(\frac{qx}{mR\Delta H}\right)\left(\frac{\Delta\rho}{\rho_{pc}}\right) \quad (20)$$

P has been graphed as a function of nondimensional axial length ($x/L_2$) for all the cases in figure 14(a). The plot shows a maximum for the first HTD location case-1, whereas, for the other cases, there is a gradual increase in the value. So this peak corresponds to the jump in the volume fraction reported in case-1. Although equation (20) is in terms of known input variables except for h that are known beforehand, a known dimensionless number in the expression is missing. Also, the expression can be made independent by reframing so that it develops the capability to predict the two-phase thickness theoretically ($h_{th}$) instead of using numerical h value ($h_{nu}$).

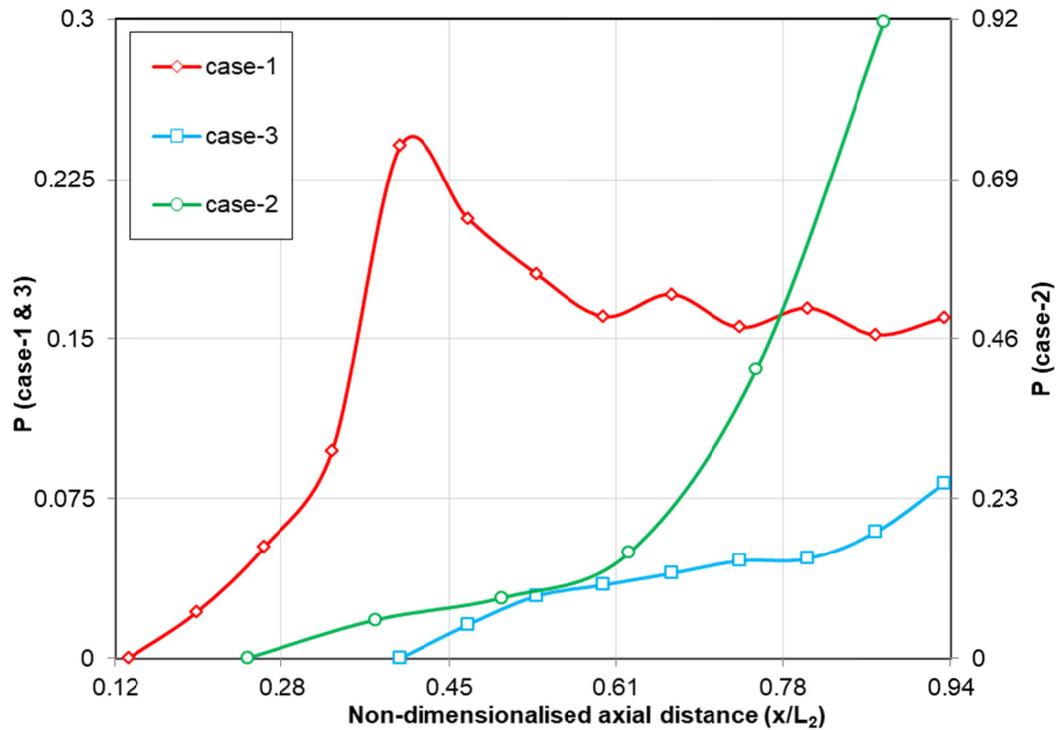

(a)

The input variables have been modified, and the following definition of them has been used to moderate the equation (20).

$$\text{Alternate definitions are} \begin{cases} q = K_e\left(\dfrac{T_w - T_{ps}}{h}\right) \\ \Delta H = C_p(T_{ps} - T_{in}) \\ m = \rho U \end{cases} \quad (21)$$

So, imbibing the above alternate definitions into equation (19) and after some adjustment, the final appearance of the equation is the following:

$$h_{th} = C(\rho^*)^{0.5}(T^*)^{0.5}(Pe_R)^{-0.5}(Rx)^{0.5} \quad (22)$$

Here $T^* = \dfrac{T_w - T_{ps}}{T_{ps} - T_{in}}$, $\rho^* = \dfrac{\Delta \rho}{\rho_{pc}}$ and Pe is a peclet number based on R. Equation shows that h will depend on $T^*$, $Pe_R$ and $(x)^{0.5}$ which will be different at any axial location in the flow domain. In the end, values of $h_{th}$ has been compared with $h_{nu}$ and the same has been plotted for C equal to 0.1 in figure 14(b). The juxtaposition reveals that equation (22) is competent enough to forecast the value of h and gives a relevant length scale for the supercritical flow.

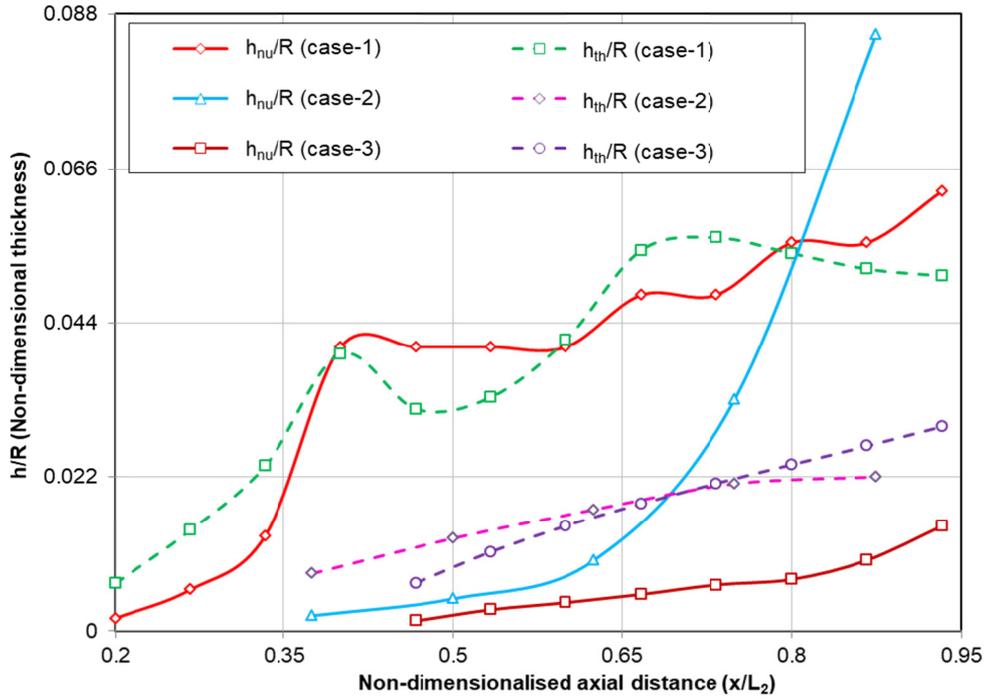

(b)

**Figure 14. (a) P variation with respect to axial location (b) Comparison of h value computed from both methods**

## 4. Conclusion

HTD in supercritical water which is reported in Shitsman's and Ornatskij's experiments, is numerically studied using an unconventional pseudo-phase change method. This is done in commercial software ANSYS FLUENT using the VOF multiphase model. The results showed that the proposed model performs similarly to the traditional single-fluid approach. The propounded quantitative analysis makes the pre-existing qualitative understanding more lucid and gives a clear picture of all the major possibilities in supercritical flows.

Further, the two-phase model equips us with an opportunity to set up an analogy with the boiling phenomena in subcritical flows. The volume fraction data shows something similar to pseudo film boiling for a particular kind of HTD, resulting from buoyancy forces. The same phenomenon is observed as the local maximum in the dimensionless two-phase thickness plot. Besides, the theoretical expression for calculating h gives a better insight into the supercritical flow, and its prediction capabilities render us an important length scale without going into much flow details.

Although the proposed method fared well, there is a scope of improvements that need to be addressed. For example, the pseudo-phase change idea employs a multiphase model that has parameters that need to be fine-tuned for supercritical fluid application. Also, the parameter that is introduced during the pseudo-phase change process, such as ΔT has to be explored. So the way forward is to conduct a lot of experiments whose results can be used to fit the model or be compared against corresponding numerical study. We can come up with the value of parameters for specific flow conditions. In the end, $h_{th}$ equation can be investigated further for the C value and how it varies for different supercritical fluids.

**Acknowledgement**

The authors will like to acknowledge the financial support from the Department of science and technology (DST), Government of India under the National Center for Clean Coal Research and development (NCCCR&D) scheme.